\documentclass[aps, reprint, superscriptaddress, showpacs, longbibliography]{revtex4-1}
\usepackage{color, graphicx}
\usepackage{amssymb,amsmath, amsfonts, bm}
\usepackage[colorlinks,citecolor=blue, urlcolor=blue, linkcolor=blue]{hyperref}
\usepackage{bbold}

\begin{document}
\title{Hidden physics in the dual-fermion approach - a special case of a non-local expansion scheme}  

\author{Gang Li}
\email[Correspondence and requests for materials should be addressed to: ]{gangli@physik.uni-wuerzburg.de}
\affiliation{\mbox{Institut f\"ur Theoretische Physik und Astrophysik,
  Universit\"at W\"urzburg, 97074 W\"urzburg, Germany}}


\begin{abstract}
  In this work, we present a non-local expansion scheme to study
  correlated electron systems aiming at a better description of its
  spatial fluctuations at all length scales.   
  Taking the non-local coupling as a perturbation to the local degrees
  of freedom, we show that the non-locality in the self-energy
  function can be efficiently constructed  from the coupling between
  local fluctuations.  
  It can provide one unified framework to incorporate non-locality to
  both ordered and disordered correlated many-body fermion systems.  
  As the first application, we prove that the dual-fermion approach
  can be understood as a special case of this non-local expansion
  scheme. The scheme presented in this work
  is constructed without introducing any dual variable, in which the
  interacting nature and the correlated behaviors of the lattice
  fermions have a clear physics correspondence.   
  Thus, in this special case, the equivalence of the dual-fermion
  approach to the non-local expansion scheme beautifully reveals the
  physics origin of the dual variables.
   We show that the non-interacting dual-fermion Green's function
  corresponds exactly to a non-local coupling of the lattice fermion
  renormalized by the local single-particle charge fluctuations, and
  the dual-fermion self-energy behaves as the one-particle fully
  irreducible components of the lattice Green's function.    
  Not only limited to this specific example, the non-local expansion scheme
  presented in this work can also be applied to other problems 
  depending on the choice of the local degrees of freedom.  
\end{abstract}

\maketitle
\section{Introduction}
The Bloch's theorem~\cite{Bloch} for electron
states in crystals is based on their translational symmetry. These
states are labeled by the quantum number $k$, which reflects their
invariant behaviours under a translation by a Bravais lattice vector.
However, the translational symmetry in a real solid
is never perfect, which is not only because of its finite dimension, but also
because of the presence of chemical potential/interaction disorders,
lattice defects, impurities and phonons, etc. The finite dimension of
a system may give rise to states which are sensitive to the surface
but insensitive to the interior of the bulk.  
The chemical potential or interaction disorder localizes electron
states  in real space which can trigger a metal-insulator transition
(also known as an Anderson transition~\cite{PhysRev.109.1492}). 
In the ultracold atom system in optical
lattices~\cite{RevModPhys.80.885}, though the lattice defects, 
impurities or phonons are absent, the spatial inhomogeneity is always
present, as harmonic confinement potential introduces a spatially
varying local density, which breaks translational symmetry of these cold
atom systems. 
In addition, in systems with short-ranged magnetic correlations, the
homogeneity can also be broken by the creation of ordered patterns in
real space.    
When the concentration of disorder or magnetic patterns is
small, the system can be treated approximatedly as if the
translational symmetry were not broken, in this case the quantum 
number $k$ can still be used to characterize the electron states.  
However, in the case with strong disorders, the translational symmetry
should be completely abandoned. 
Thus, it is important to have a unified theoretical framework to
capture the effects of spatial inhomogeneity, espectially in the
presence of strong electronic correlations.   
 
This problem represents a strong challenge to modern many-body
theories.  
Due to the presence of the strong electronic correlation, approaches
formulated from either weak-coupling or strong-coupling expansions,
such as fluctuation exchange (FLEX)~\cite{PhysRevLett.62.961}, random
phase approximation (RPA), cumulant expansion~\cite{0022-3719-21-18-002, PhysRevB.43.8549}, strong-coupling expansion~\cite{doi:10.1142/S0217979200002430, PhysRevLett.80.5389}. 
etc. are not adequate to handle the complete parameter range in
interaction.  
Unbiased numerical approaches, such as exact-diagonalization and
quantum Monte Carlo~\cite{PhysRevD.24.2278, PhysRevB.24.4295}, can treat
the correlation effect precisely, but with 
the penalty on their finite cluster size, {\it i.e.}
the thermodynamic limit is not directly available in these approaches. 
Another class of many-body approaches, based on the dynamical
mean-field theory (DMFT)~\cite{RevModPhys.68.13}, can fairly treat
electronic correlation at arbitrary strength and contain the
thermodynamic limit from their construction.
It has been shown that, the DMFT can provide very important insights
for several nonperturbative properties, such as the Mott-Hubbard
transition. It also shows its great power in the study of correlated
inhomogeneous systems. 
In the so-called real-space DMFT
(R-DMFT)~\cite{PhysRevLett.100.056403, 1751-8121-44-49-495004}, the
correlated and disordered sites are treated as a group of Anderson
impurities, which are embedded into a lattice system at the
thermodynamic limit via the DMFT self-consistency. 
However, the DMFT is exact only in the infinite
spatial dimension limit, where the character of correlation effects is
purely local in space. 
For realistic systems at finite dimension, the DMFT downgrades to an
approximation, which neglects the spatial fluctuation effect beyond the
mean-field level.
For this reason, in the R-DMFT, disorder and correlation effect is
treated locally. The spatial fluctuations of the complete 
system are determined by the coupling of the impurities within
strong-coupling perturbation theory at leading order only (mean-field
level).    
Generalizing an impurity to a cluster incorporates the short-ranged
spatial fluctuations inside the cluster. Both, a reciprocal-space
(dynamical cluster approximation, DCA~\cite{RevModPhys.77.1027,
  PhysRevB.61.12739, PhysRevB.64.195130, PhysRevB.61.7887, PhysRevLett.87.167010}) and a real-space construction
(cellular dynamical mean field theory,
Cellular-DMFT~\cite{PhysRevB.67.075110, PhysRevLett.87.186401}) have been suggested.  
These approaches improve results for $D=1,2,3$,  while the longer-ranged spatial
fluctuations beyond the mean-field level are still missing.  
Alternative to the cluster extensions of the DMFT, the non-local fluctuations can also be restored by the so-called ``diagrammatic expansion" methods. 
These methods include the ``advanced" fluctuation-exchange scheme~\cite{doi:10.1143/JPSJ.75.054713}, dynamical vertex approximation (D$\Gamma$A)~\cite{PhysRevB.75.045118, Held01062008}, the dual-fermion (DF) approach~\cite{PhysRevB.77.033101, PhysRevB.77.195105, PhysRevLett.102.206401} and the one-particle irreducible functional approach~\cite{PhysRevB.88.115112}.
The non-locality in these methods is constructed from the dynamical scattering of multiple particles at the same spatial location. 
Comparing to the cluster extensions of the DMFT, it can be shown that both short- and long-range fluctuations are equally treated in these methods.
The D$\Gamma$A and the DF can also be easily adapted to the study of inhomogeneous systems~\cite{PhysRevLett.104.246402, PhysRevB.89.195116} and shows their advantage over the R-DMFT.   

In this paper, in line with all other ``diagrammatic expansion" methods, we consider a correlated system without translational
symmetry and aim at a better description of the spatial fluctuations at
all length scales in the presence of strong electron-electron
interaction. 
To this purpose, we generalize the strong-coupling
expansion approach (cumulant expansion) to such a system and treat the
non-local coupling as perturbations. The expansion of the non-local
coupling can generate, order by order, the non-local spatial
fluctuations from the coupling of the local charge fluctuations at
different spatial locations.  
Compared to the other ``diagrammatic expansion" methods mentioned above, the derivation of this scheme is rather general and the choice of the local system is quite flexible, which may provide valuable insights and understanding into other methods. 
For example, as will be shown in this paper, the DF approach can be understood as one special case of this non-local expansion scheme.  
Furthermore, depending on the choice of the local system, this scheme can either
nicely go beyond the R-DMFT approximation to result in a non-local
self-energy for correlated disordered systems or provide a quick
cluster solver for the Cellular-DMFT. Not limited to these two specific
applications, our scheme provides one unified framework to incorporate
non-locality to a correlated many-body system with only moderate
numerical cost. 

\section{Non-local expansion}\label{formalism}
The translational symmetry of a homogeneous correlated
system can be broken, for example, by disorders in chemical
potential or interactions. 
But inhomogeneity in a real system can be much more general and
have many different origins.
The only assumption we make in this work is that the interaction shall
be local. An example Hamiltonian looks like the following,
\begin{equation}\label{hamiltonian}
H=-\sum_{i,j} (c_{i\sigma}^{\dagger}\mathbf{t}_{i,j}c_{j\sigma}+h.c.)
- \sum_{i}\mu_{i}n_{i} + \sum_{i}U_{i}n_{i\uparrow}n_{i\downarrow}\:.
\end{equation}
Here, the values of the chemical potential $\mu_{i}$ and the Coulomb
repulsion $U_{i}$ depend on their spatial
coordinations, their different values in space give rise to the
spatial inhomogeneity. $\mathbf{t}_{ij}$, in this equation, does not
have to be restricted to only nearest neighbors. It
can be quite general to contain more hopping terms, or spatial
anisotropy, etc.   
Eq.~(\ref{hamiltonian}) represents one type of
inhomogeneity that the systems could have.
As we discussed before, there are many other ways to induce
disorders to the systems. Not losing generality, we discard
the specific form of the Hamiltonian and only cast it into two parts
for the convenience of the following discussions.
\begin{equation}
H = \sum_{i=1}^{N}H_{i} + H^{NL}\:.
\end{equation}
$H_{i}$ gathers every term that is locally related to site $i$. All
other terms that carry non-locality are grouped into $H^{NL}$. 
The corresponding action can be written as
\begin{eqnarray}\label{full-action}
{\cal S}&=&\int{\mbox d}\tau{\mbox d}\tau^{\prime}\sum_{i,j}c_{i\sigma}^{*}(\tau)[{\cal
    G}^{-1}(\tau-\tau^{\prime})]_{i,j}c_{j\sigma}(\tau^{\prime})\nonumber\\
&&+\sum_{i}\int{\mbox
d}\tau U_{i}n_{i\uparrow}(\tau)n_{i\downarrow}(\tau)\nonumber\\
&=&\sum_{i=1}^{N}{\cal S}_{i}+\int{\mbox d}\tau{\mbox
  d}\tau^{\prime}\sum_{i\ne j}c_{i\sigma}^{*}(\tau)[{\cal
    G}^{-1}_{\sigma}(\tau-\tau^{\prime})]_{i,j}c_{j\sigma}(\tau^{\prime})\nonumber\\
&=&\sum_{i=1}^{N}{\cal S}_{i}
+T\sum_{i\ne
  j}\sum_{\omega}c_{i\omega\sigma}^{*}[{\cal G}^{-1}_{\sigma}(\omega)]_{ij}c_{j\omega\sigma}   
\end{eqnarray}
where $[{\cal G}(\omega)_{\sigma}^{-1}]_{ij}$ describes the dynamic
coupling between states at different spacial locations. 
${\cal S}_{i}$ is the action that contains only the local
degrees of freedom, in which $[{\cal G}^{-1}_{\sigma}(\omega)]_{ii}$
can have many different forms. For example, in the cellular-DMFT, $[{\cal
    G}(\omega)_{\sigma}^{-1}]_{ii}=-(\omega + \mu) +
\Delta_{ii}(\omega)$, which is the local component of the inverse
Weiss field. $[{\cal G}(\omega)_{\sigma}^{-1}]_{i\ne j}$ is the
non-local hybridization function $\Delta_{i\ne j}(\omega)$. 
There are also many other possibilities of $[{\cal
    G}(\omega)_{\sigma}^{-1}]_{i\ne j}$. For the moment we keep $[{\cal
    G}(\omega)_{\sigma}^{-1}]_{i\ne j}$ as 
a general function that is purely non-local in space. At the end of
this section, we will examine some specific forms of $[{\cal
    G}(\omega)_{\sigma}^{-1}]_{ij}$ to show that, actually, the
formalism presented in this work can be nicely linked to some widely-known powerful methods, especially it can extend these local many-body approaches to partially include non-trivial spatial fluctuations. 
To be more precise, we refer the non-trivial spatial fluctuations to
the spatial dependence in the self-energy function. If self-energy
is purely a local function, the corresponding spatial fluctuations are
trivial. 

\subsection{diagram expansion}
The general idea of the non-local expansion approach is very
simple:  we take the second term
on the r.h.s. of Eq.~(\ref{full-action}) as a perturbation to the
local action ${\cal S}_{i}$. 
By expanding this term, we will be able to determine the
non-local dynamic quantity, such as the single-particle Green's
function, from the local action ${\cal S}_{i}$. For simplicity, we
take $[{\cal G}(\omega)_{\sigma}^{-1}]_{ij}$ as $V_{ij}^{\omega}$ and 
$\omega\equiv(\omega,\sigma)$.   
With Eq.~(\ref{full-action}), the single-particle
Green's function is calculated as 
\begin{widetext}
\begin{eqnarray}\label{G-expansion}
G_{\alpha\beta}^{\omega}&=&-\langle 
c_{\alpha\omega}c_{\beta\omega}^{*}\rangle
=-\frac{1}{\cal Z}\int {\cal
  D}[c^{*}, c] \exp\left[-\sum_{i=1}^{N}{\cal
    S}[c_{i}^{*},c_{i}] - T\sum_{i\ne
  j}\sum_{\omega^{\prime}}c_{i\omega^{\prime}}^{*}V_{ij}^{\omega^{\prime}}c_{j\omega^{\prime}}\right]c_{\alpha\omega}c_{\beta\omega}^{*}
\nonumber\\
&=&-\frac{1}{\cal Z}\prod_{i=1}^{N}
\int {\cal D}[c_{i}^{*}, c_{i}]e^{-{\cal S}_{i}[c_{i}^{*},c_{i}]}
\sum_{n=0}^{\infty}\frac{(-T)^{n}}{n!}
\left[\sum_{i\ne
j}\sum_{\omega^{\prime}}c_{i\omega^{\prime}}^{*}V_{ij}^{\omega^{\prime}}c_{j\omega^{\prime}}\right]^{n}
c_{\alpha\omega}c_{\beta\omega}^{*}\:.
\end{eqnarray} 
\end{widetext}
Here, $\alpha,\beta$ are two arbitrary spatial indices. 
${\cal Z}$ is the full partition function containing both local and non-local contributions. 
\begin{equation}\label{partition-c}
{\cal Z} = \int {\cal
  D}[c^{*}, c] \exp\left[-{\cal S}_{i}[c_{i}^{*},c_{i}] - T\sum_{i\ne
  j}\sum_{\omega^{\prime}}c_{i\omega^{\prime}}^{*}V_{ij}^{\omega^{\prime}}
  c_{j\omega^{\prime}}\right]\:. 
\end{equation}

${\cal Z}$ can be replaced by the local partition function ${\cal
  Z}_{loc}=\prod_{i=1}^{N}{\cal Z}_{i}$ when only the connected
diagrams are considered in Eq.~(\ref{G-expansion}), as we will do
in this work.  
In a calculation with only the local ${\cal S}_{i}$, the
single-particle Green's function one can get is the one with
$\alpha=\beta$, 
\begin{equation}\label{g-DMFT}
g_{\alpha}^{\omega}=-\prod_{i=1}^{N}\frac{1}{{\cal Z}_{i}}\int {\cal
  D}[c^{*}_{i}, c_{i}] \exp\left[-{\cal
    S}_{i}[c^{*}_{i},c_{i}]\right]c_{\alpha\omega}c_{\alpha\omega}^{*}\:.
\end{equation} 
All terms for $\alpha\ne\beta$ vanish as the two operators, $c_{\omega\alpha}$
and $c_{\omega\beta}^{*}$, must pair in the above grassmann integral. 
With this in mind, we will proceed to evaluate the expansion in 
Eq.~(\ref{G-expansion}) order by order. 
Before proceeding, we want to note that Eq.~(\ref{G-expansion}) was
more often evaluated by introducting a dual variable from the  
Hubbard-Stratonovich transformation~\cite{PhysRevLett.3.77, 1957SPhD416S} on the non-local term. 
Here, we want to keep  the present form of Eq.~(\ref{G-expansion})
and do not introduce any dual variable.  At the end of this
paper, we will compare our formalism to those obtained from
the Hubbard-Stratonovich transformation to illuminate the physics
hidden behind the introduction of the dual variables in those
approaches.  

\begin{figure}[htbp]
\centering
\includegraphics[width=\linewidth]{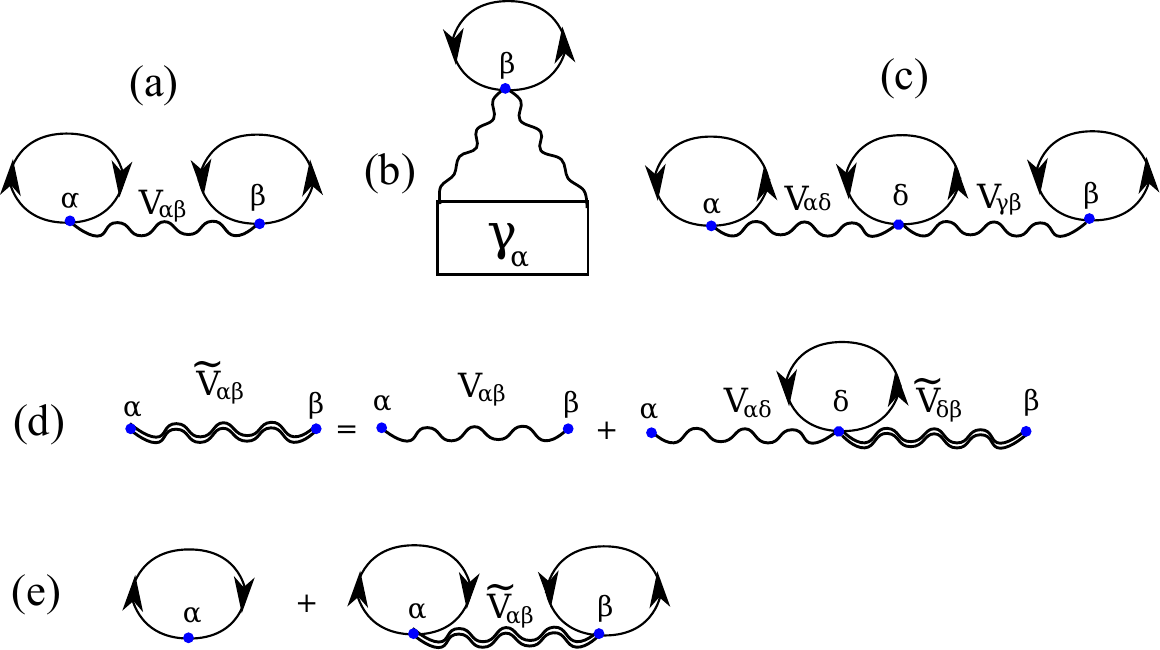}
\caption{A few lowest order Feynmann diagrams in the expansion of
  Eq.~(\ref{G-expansion}). (a), (b) and (c) correspond to the first
three expansions. (d) The local single-particle charge fluctuations
renormalize the non-local hybridization. With the renormalized hybridization $\tilde{V}_{\alpha\beta}$, diagrams similar to (a) and (c) now can be simply written as (e). }
\label{Fig1}
\end{figure}

The first term in the expansion of Eq.~(\ref{G-expansion}) is exactly
the local Green's function as defined in Eq.~(\ref{g-DMFT}). In the
second term, $\alpha$ and $\beta$ must be different as $i\ne j$. The
only way to pair these operators is to set $i=\alpha$ and $\beta=j$,
which gives rise to the first {\it non-local} correction to ${\cal
  S}_{i}$, 
\begin{equation}\label{g-NL-a}
G_{\alpha\ne\beta}^{\omega,(a)}=g_{\alpha}^{\omega}V_{\alpha\beta}^{\omega}g_{\beta}^{\omega}\: .
\end{equation}
The graphical representation of this term is shown in
Fig.~\ref{Fig1}(a). The wiggly line represents the non-local
hybridization $V_{\alpha\beta}$ between the two spatial locations
$\alpha$ and $\beta$. The 
solid line circling at each location represents the local Green's
function computed from ${\cal S}_{i}$, as given in Eq.~(\ref{g-DMFT}).

The third expansion term in Eq.~(\ref{G-expansion}) contains two
topologically distinct diagrams.
They correspond to the corrections of the non-local charge
fluctuations to the local and non-local Green's functions,
respectively.  
When $\alpha=\beta$, this term gives the first correction to the
local Green's function $g_{\alpha}^{\omega}$, which is generated from
the non-local charge fluctuations coupled with the local
two-particle vertex. To see this, we need to examine the possible
contractions of operators connecting to
$c_{\alpha\omega}c_{\alpha\omega}^{*}$. 
In order to have a connected diagram, one has to link either
$c_{j\omega^{\prime}}c_{k\omega^{\prime\prime}}^{*}$ or
$c_{i\omega^{\prime}}^{*}c_{l\omega^{\prime\prime}}$ to
$c_{\alpha\omega}c_{\alpha\omega}^{*}$ with the corresponding constraint
$j=k=\alpha$ or $i=l=\alpha$. 
We notice that the interchange of indices $i,j$ with $k,l$ does
not generate topologically new diagrams, thus, we can focus on only
one of them and neglect the factor $1/2$, {\it e.g.} $j=k=\alpha$.
\begin{widetext}
\begin{eqnarray}\label{g-NL-b}
G_{\alpha=\beta}^{\omega,(b)}&=&\frac{T^{2}}{{\cal Z}_{loc}}
\sum_{i\ne j}\sum_{k\ne l}\sum_{\omega^{\prime},\omega^{\prime\prime}}
\int {\cal D}[c^{*}, c] e^{-{\cal
    S}_{loc}[c^{*},c]}
c_{l\omega^{\prime\prime}}c_{i\omega^{\prime}}^{*}V_{i,j}^{\omega^{\prime}}V_{k,l}^{\omega^{\prime\prime}}
c_{j\omega^\prime}c_{k\omega^{\prime\prime}}^{*}c_{\alpha\omega}c_{\alpha\omega}^{*}
\delta_{\omega^{\prime},\omega^{\prime\prime}}\delta_{i,l}\delta_{j,k}
\nonumber \\
&=& T\sum_{i}\sum_{\omega^{\prime}}\langle
c_{i\omega^{\prime\prime}}c_{i\omega^{\prime}}^{*}\rangle V_{i,\alpha}^{\omega^{\prime}}V_{\alpha,i}^{\omega^{\prime\prime}}
\langle
c_{\alpha\omega^\prime}c_{\alpha\omega^{\prime\prime}}^{*}c_{\alpha\omega}c_{\alpha\omega}^{*}
\rangle 
\delta_{\omega^{\prime},\omega^{\prime\prime}}
\:.
\end{eqnarray}
The last term in the above equation is the local four-point correlation
function computed from ${\cal S}_{i}$.
\begin{equation}\label{four-point}
\langle
c_{\alpha\omega^\prime}c_{\alpha\omega^{\prime\prime}}^{*}c_{\alpha\omega}c_{\alpha\omega}^{*}\rangle
= \langle
c_{\alpha\omega^\prime}c_{\alpha\omega^{\prime\prime}}^{*}c_{\alpha\omega}c_{\alpha\omega}^{*}\rangle_{c}
+ \langle
c_{\alpha\omega^\prime}c_{\alpha\omega^{\prime\prime}}^{*}\rangle\langle
c_{\alpha\omega}c_{\alpha\omega}^{*}\rangle - 
\langle
c_{\alpha\omega^\prime}c_{\alpha\omega}^{*}\rangle\langle
c_{\alpha\omega}c_{\alpha\omega^{\prime\prime}}^{*}\rangle\:,
\end{equation}
which contains both connected and disconnected parts. 
Eq.~(\ref{g-NL-b}) corresponds to Fig.~\ref{Fig1}(b) if the connected
four-point correlation function is used. The connected four-point
correlation function $\langle c_{\alpha\omega^\prime}c_{\alpha\omega^{\prime\prime}}^{*}c_{\alpha\omega}c_{\alpha\omega}^{*}\rangle_{c}$ (after decided by four single-particle Green's function ) is also called the complete two-particle vertex,
which is graphically represented as the square in Fig.~\ref{Fig1}(b). 
 
The disconnected part of $\langle
c_{\alpha\omega^\prime}c_{\alpha\omega^{\prime\prime}}^{*}c_{\alpha\omega}c_{\alpha\omega}^{*}\rangle$
gives rise to a diagram topologically equivalent to Fig.~\ref{Fig1}(c) but
with $\alpha=\beta$. This is easier to see by 
first assuming $\alpha\ne\beta$ in the third order expansion, and
releasing this constraint at the end of the discussion,
\begin{eqnarray}\label{g-NL-c}
G_{\alpha\ne\beta}^{\omega,(c)}&=&-\frac{T^{2}}{{\cal Z}_{loc}}
\sum_{i\ne j}\sum_{k\ne l}\sum_{\omega^{\prime},\omega^{\prime\prime}}
\int {\cal D}[c^{*}, c] e^{-{\cal S}_{loc}[c^{*},c]}
c_{\alpha\omega}c_{i\omega^{\prime}}^{*}V_{i,j}^{\omega^{\prime}}V_{k,l}^{\omega^{\prime\prime}}
c_{j\omega^\prime}c_{k\omega^{\prime\prime}}^{*}c_{l\omega^{\prime\prime}}c_{\beta\omega}^{*}
\nonumber \\
&=&-\sum_{j}\langle
c_{\alpha\omega}c_{i\omega^{\prime}}^{*}\rangle V_{i,j}^{\omega^{\prime}}V_{k,l}^{\omega^{\prime\prime}}
\langle c_{j\omega^\prime}c_{k\omega^{\prime\prime}}^{*}\rangle
\langle c_{l\omega^{\prime\prime}}c_{\beta\omega}^{*}\rangle
\delta_{i,\alpha}\delta_{j,k}\delta_{l,\beta}\delta_{\omega,\omega^{\prime}}\delta_{\omega^{\prime},\omega^{\prime\prime}}
=g_{\alpha}^{\omega}\left(\sum_{j}V_{\alpha,j}^{\omega}g_{j}^{\omega}
V_{j,\beta}^{\omega}
\right)g_{\beta}^{\omega}\:.
\end{eqnarray}
\end{widetext}
Similar to Fig.~\ref{Fig1}(a), this diagram consists of only local
single-particle charge fluctuations at different spatial locations
which are connected by their dynamic hybridization. 
There will be more similar diagrams in the higher order expansions of
Eq.~(\ref{G-expansion}).  
These terms can be cast into a more compact form by renormalizing the
hybridization as shown in Fig.~\ref{Fig1}(d). The dressed hybridization
is simply found to be 
\begin{equation}\label{dressed_hybridization}
\tilde{V}_{\alpha\beta} = [V^{-1} - g\mathbb{1}]^{-1}_{\alpha\beta}\:,
\end{equation}
where $\mathbb{1}$ is an unit matrix. By replacing the
single wiggly line with the double 
wiggly line, Fig.~\ref{Fig1}(a) becomes Fig.~\ref{Fig1}(e), which now
incorporates all higher order diagrams similar as
Fig.~\ref{Fig1}(c). 
With this substitution, we can now release the constraint of $\alpha\ne\beta$.
Setting $\alpha=\beta$ in Fig.~\ref{Fig1}(c) does not violate any feynmann
rule we used above, and it corresponds exactly to the diagram with the
disconnected four-point correlation function in
Eq.~(\ref{four-point}). In fact, one can easily notice that only the 
crossing term, {\it i.e.} the last term of Eq.~(\ref{four-point}), leads 
to a connected diagram of $G_{\alpha\beta}^{\omega}$, which is same as what we derived in 
Eq.~(\ref{g-NL-c}), but with $\alpha=\beta$.  

Repeating the above procedure, one could, in principle, calculate all expansion terms, which contain the local complete vertices at one-, two-, $\cdots$, n-particle levels.  
The one-particle Green's function calculated in Eq.~(\ref{G-expansion}) is then dressed by these local scattering processes. 
It is easy to see that the non-locality is generated from the non-local coupling of these local scattering modes. 
The more terms are considered in the expansion, the better approximation of $G_{\alpha\beta}^{\omega}$  
will be obtained. 
However, for practical reason, it is not possible to consider all expansion terms. A truncation of the expansion has to be done. 
As all expansion terms contribute to the calculation of $G_{\alpha\beta}^{\omega}$ in Eq.~(\ref{G-expansion}), there is no principle to favour certain terms over the others in the expansion.
As we will see in the next section, however, a simple truncation of the expansion at certain order will violate some exact limit of the problem. 
Only by summing certain similar diagrams up to the infinite order, one can reproduce these limits. 
In the next section, we will examine one truncation scheme (more precisely, one diagram resummation scheme), which respects these exact limits and is feasible for practical implementation.   

\subsection{diagram resummation}
First, we show that the inclusion of all diagrams similar to Fig.~\ref{Fig1}(a) and (c) is
of crucial importance to our non-local expansion scheme.
It has a very crucial implification in the non-interacting limit. With
the dressed hybridization in Eq.~(\ref{dressed_hybridization}),
Fig.~\ref{Fig1}(a), (c) and all other similar diagrams can be cast into
one simple diagram shown as Fig.~\ref{Fig1}(e). The single-particle
Green's function evaluated from this diagram is given as:
\begin{equation}\label{g-NL-e}
G_{\alpha\beta}^{\omega, (e)}=g_{\alpha}^{\omega}\delta_{\alpha,\beta}+g_{\alpha}^{\omega}\tilde{V}_{\alpha\beta}^{\omega}g_{\beta}^{\omega}\:.
\end{equation}
In the following, we will show that, with the dressed hybridization,
Eq.~(\ref{g-NL-e}) gives rise to $G_{\alpha\beta}^{0,\omega}$ exactly. 
This is a crucial condition to fulfill, as the non-interacting limit is an
intuitive case to verify a diagrammatic approach. 
When interaction is switched off, {\it i.e.} $U=0$, all
reducible multi-particle susceptibilities vanish, thus,
the only diagram we need to evaluate is  Fig.~\ref{Fig1}(e). 
\begin{equation}
G_{\alpha\beta}^{\omega} =
g_{\alpha}^{0,\omega}\delta_{\alpha\beta} +
g_{\alpha\alpha}^{0,\omega}[V^{-1}-g^{0}\mathbb{1}]^{-1}_{\alpha\beta}g^{0,\omega}_{\beta}\:. 
\end{equation}
In this equation, $g_{\alpha}^{0}$ is the local single-particle Green's
function in the non-interacting limit, which shall be understood as
$[[{\cal G}_{\sigma}(\omega)]_{\alpha\alpha}^{-1}]^{-1}$. It is different from
the local Weiss field ${\cal G}_{\alpha\alpha}(\omega)$.
\begin{eqnarray}\label{G0}
G_{\alpha\beta}^{\omega} &=&
g_{\alpha}^{0,\omega}\delta_{\alpha\beta} +
g_{\alpha}^{0,\omega}\left[\frac{V g^{0}}{\mathbb{1} -
    V g^{0}}\right]_{\alpha\beta}\nonumber\\
&=&g_{\alpha}^{0,\omega}\delta_{\alpha\beta} -
g_{\alpha}^{0,\omega}\left[
\mathbb{1}-\frac{\mathbb{1}}{\mathbb{1}-V
  g^{0}}\right]_{\alpha\beta}\nonumber\\
&=&\left[\frac{\mathbb{1}}{g^{0,-1}\mathbb{1}-V}\right]_{\alpha\beta}
={\cal G}^{\omega}_{\alpha\beta}\:.
\end{eqnarray}
Thus, the inclusion of local one-particle charge fluctuations at
different spatial locations under the coupling of non-local
hybridizations gives rise to the exact dynamics at all length scales in
the non-interacting limit. 
At this point, it shall also be mentioned that, the current scheme can
also correctly describe the strong coupling limit. In the strong
coupling limit, the local dynamics become dominant, the expansion
over $V_{\alpha\beta}$ becomes less important. ${\cal S}_{loc}$
accounts for the major dynamics in that case. 

\begin{figure}[htbp]
\centering
\includegraphics[width=0.9\linewidth]{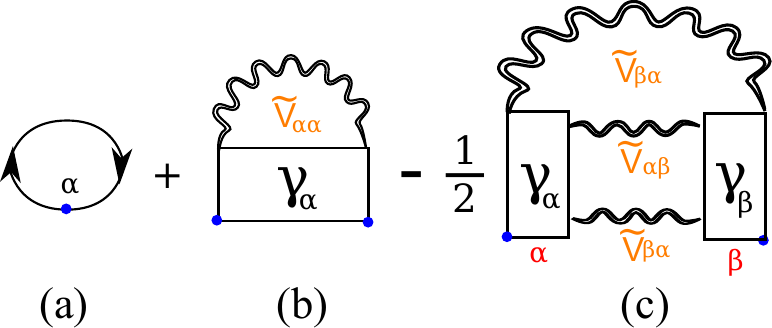}
\caption{The one-particle irreducible diagrams
  $\Lambda_{\alpha\beta}^{\omega}$ used for constructing the connected
  lattice Green's function in the non-local expansion scheme.} 
\label{Fig2}
\end{figure}
In the intermediate coupling region, the diagrams with vertices
become relevant. One example is Fig.~\ref{Fig1}(b), and more similar
diagrams appear in the higher order expansions of
Eq.~(\ref{G-expansion}).   
In these diagrams, similar to the renormalization of the
hybridization shown in Eq.~(\ref{dressed_hybridization}), the higher
order terms can be viewed as the dressed diagrams of lower order
terms. 
Thus, we can cast these diagrams into a more compact form by using these
lower-order diagrams as building blocks.
Certainly, there are different ways to construct the higher-order diagrams from
these building blocks, which corresponds to the different choices of the
subset of the complete Feynmann diagrams in the expansion of
Eq.~(\ref{G-expansion}).  
In this work, we consider the following
resummation scheme whose physics meaning becomes clear in the end of this
section.   
\begin{equation}\label{g-NL}
G^{\omega}_{\alpha\beta} = \left[\frac{\Lambda^{\omega}}{\mathbb{1}-V^{\omega}\Lambda^{\omega}}\right]_{\alpha\beta}\:,
\end{equation}  
in which $\Lambda^{\omega}_{\alpha\beta}$ is the one-particle fully
irreducible (1PI) diagrams taken as the construction blocks.  
Eq.~(\ref{g-NL}) connects different 1PI diagrams of the
single-particle propagator by the bare hybridization
$V_{\alpha\beta}^{\omega}$.   
Intuitively, in Eq.~(\ref{g-NL}), if we take the simplest 1PI
diagram, which is $g^{\omega}_{\alpha\beta}\delta_{\alpha\beta}$,
Eq.~(\ref{g-NL-e}) is reproduced. 
It is obvious that, the more 1PI diagrams are adopted in
$\Lambda_{\alpha\beta}^{\omega}$, the better approximation we can get
for $G^{\omega}_{\alpha\beta}$. 
As one will see in the next section, Eq.~(\ref{g-NL}) can reproduce
one advanced many-body approach, {\it i.e.} dual-fermion approach~\cite{PhysRevB.77.033101} ,
when the local system is taken as the DMFT impurity.  

In these building block diagrams, not only the two-particle vertices but also higher-order vertices will appear. 
For practical reasons, we restrict our calculations to the two-particle vertices. 
It shall be noted here that, in addition to Eq.~(\ref{g-NL}), one can also formulate other ways of summing diagrams with  vertices. 
For simplicity, in this work we consider the 1PI diagrams 
Fig.~\ref{Fig2} for $\Lambda^{\omega}_{\alpha\beta}$. 
The first diagram in Fig.~\ref{Fig2} is the local single particle propagator
$g^{\omega}_{\alpha\beta}\delta_{\alpha\beta}$ calculated from the
local ${\cal S}_{\alpha}$. 
This diagram only gives rise to a mean-field description
of the inter-site Green's function, which equivalently means that the self-energy
computed from this diagram is still local.
This can be seen by starting from Eq.~(\ref{g-NL-e}) and repeating the
derivation in Eq.~(\ref{G0}) with
$g_{\alpha}^{-1}=g_{\alpha}^{0,-1}-\Sigma_{\alpha}$. One can easily
see that the lattice self-energy function
$\Sigma_{\alpha,\beta}=\Sigma_{\alpha}\delta_{\alpha,\beta}$ is
exactly the same as the local self-energy obtained from ${\cal
  S}_{\alpha}$, {\it i.e.} no non-locality is included in the self-energy.
Fig.~\ref{Fig2}(b) corresponds to the diagrams
with single-particle charge fluctuations coupled to local
two-particle scattering modes, which is obtained from
Fig.~\ref{Fig1}(b) by replacing the hybridization with the dressed
one, see Eq.~(\ref{dressed_hybridization}).
It is crucial to note that, though this diagram is local in space,
through Eq.~(\ref{g-NL}), it generates spatial dependence
of the {\it self-energy function} that is missing in
Fig.~\ref{Fig2}(a).   
Fig.~\ref{Fig2}(c) consists of two connected
four-point correlation functions
$\gamma_{\alpha}^{\sigma_{1}\sigma_{2};\sigma_{3}\sigma_{4}}=\langle c_{\alpha\omega_{1}}c_{\alpha\omega_{2}}^{*}c_{\alpha\omega_{3}}c_{\alpha\omega_{4}}^{*}\rangle_{c}$.
In addition to the second diagram, it generates further non-local
corrections to the lattice self-energy. 
The sum of the internal spin indices involved in this diagram can be more
conveniently carried out by introducing
$\gamma^{c/s}_{\alpha}=\gamma^{\uparrow\uparrow\uparrow\uparrow}_{\alpha}\pm\gamma^{\uparrow\uparrow\downarrow\downarrow}_{\alpha}$. 
$\Lambda_{\alpha\beta}^{\omega}$ is then found to be: 
\begin{eqnarray}\label{g-NL-L1}
&&\Lambda_{\alpha\beta}^{\omega} = g^{\omega}_{\alpha\beta}\delta_{\alpha\beta}-T\sum_{2}\gamma_{\alpha}^{\sigma_{1}\sigma_{1};\sigma_{2}\sigma_{2}}(11;22)\tilde{V}_{\alpha\alpha}(\omega_{2})\nonumber\\
&&\hspace{0.8cm}-\frac{T^{2}}{4}\sum_{234}\tilde{V}_{\beta\alpha}(\omega_{2})
\tilde{V}_{\beta\alpha}(\omega_{4})
\tilde{V}_{\alpha\beta}(\omega_{3})
\nonumber\\
&&\hspace{0.8cm}\left[
3\gamma_{\alpha}^{s}(12;34)\gamma^{s}_{\beta}(43;21)+\gamma_{\alpha}^{c}(12;34)\gamma_{\beta}^{c}(43;21)
\right]\:.
\end{eqnarray}
 Here a compact form of frequency index is used, $1\equiv(\omega_{1},\sigma_{1})$. 
 The energy conservation requires $\omega_{1}+\omega_{3}=\omega_{2}+\omega_{4}$ in the above equation.
 
Eq.~(\ref{g-NL}) and Eq.~(\ref{g-NL-L1}) are the main results of our
non-local expansion scheme.  
To construct non-locality from Eq.~(\ref{g-NL-L1}), only the local
single-particle Green's function $g_{\alpha}^{\omega}$ and the
connected four-point correlation function
$\gamma_{\alpha}^{c/s}(12;34)$ are required. They are obtained from
the solution of the local problem defined by ${\cal 
  S}_{\alpha}$, which can be solved with modern numerical
algorithms. Thus, the total number of correlated sites $N$ in
these equations can be reasonably large to represent the
thermodynamic limit. In this sense, both the short- and the long-range
spatial fluctuations can be described by this formalism. 

The computation flow of this new scheme is the following: one starts
with the construction of the Weiss field ${\cal G}^{\omega}_{\alpha\beta}$
and separates the local and non-local components as shown in
Eq.~(\ref{full-action}).
The local problem defined by the local component ${\cal
  G}^{\omega}_{\alpha\alpha}$ and $\mu_{\alpha}$, $U_{\alpha}$ is then
solved to get the local single-particle Green's function
$g_{\alpha}^{\omega}$ and the connected four-point correlation
function $\gamma_{\alpha}^{c/s}(12;34)$.  
Now Eq.~(\ref{g-NL-L1}) can be issued to
compute the Green's functions $\Lambda_{\alpha\beta}^{\omega}$, which
contain the non-trivial spatial fluctuations.  
With the new Green's function $G_{\alpha\beta}^{\omega}$ from
Eq.~(\ref{g-NL}), the self-energy $\Sigma_{\alpha\beta}^{\omega}$ can
be calculated from Dyson equation, which contains both the local and
non-local components and serves as the input for the next iteration of
this calculation. The entire calculation shall stop only when
$\Sigma_{\alpha\beta}^{\omega}$ does not change anymore. 

The current scheme is formulated for the correlated system without
translational symmetry. But it also applies to systems
characterized by the quantum number $k$. As one unified scheme for
both cases, here we also present the corresponding formula for the
latter case. When the system is translationally invariant, only one
impurity problem needs to be solved. The local impurity correlation
functions $g^{\omega}$ and $\gamma^{s/c}$ become completely site
independent. After applying Fourier transformation to Eq.~(\ref{g-NL}) and
Eq.~(\ref{g-NL-L1}), we have the following expressions for the lattice  
Green's function from the non-local expansion scheme: 
\begin{subequations}\label{g-NL-k}
\begin{align}
&G_{k}^{\omega} = \Lambda_{k}^{\omega}/[1-V_{k}^{\omega}\Lambda_{k}^{\omega}]\:, \\
&\Lambda_{k}^{\omega}=g^{\omega}-T\sum_{2}\gamma_{\alpha}^{\sigma_{1}\sigma_{1};\sigma_{2}\sigma_{2}}(11;22)\tilde{V}_{\alpha\alpha}(\omega_{2})\nonumber\\
&\hspace{0.3cm}-\frac{T^{2}}{4N^{2}}\sum_{234}\sum_{k^{\prime},q}\tilde{V}_{k+q}(\omega_{2})\tilde{V}_{k^{\prime}}(\omega_{4})\tilde{V}_{k^{\prime}+q}(\omega_{3})\nonumber\\
&\hspace{0.3cm}\left[3\gamma^{s}(12;34)\gamma^{s}(43;21)+\gamma^{c}(12;34)\gamma^{c}(43;21)\right]\:.
\end{align}
\end{subequations}

\begin{figure}[htbp]
\centering
\includegraphics[width=\linewidth]{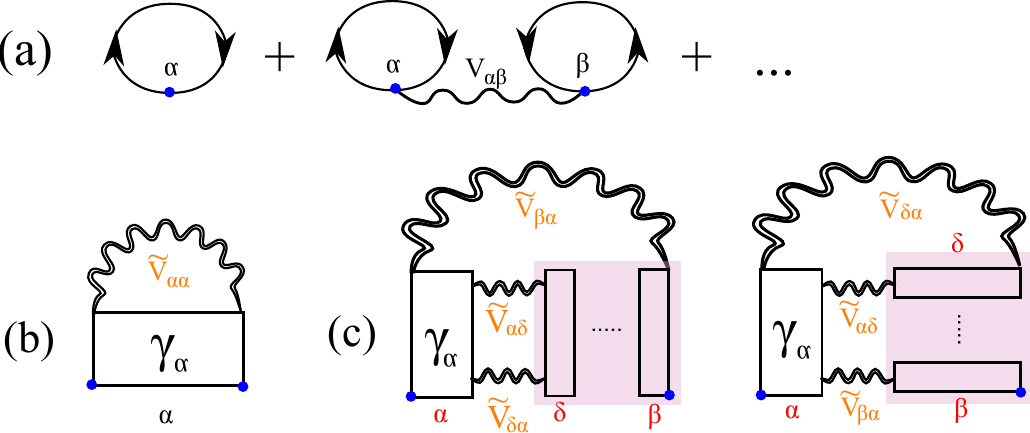}
\caption{Another set of building-block diagrams and the corresponding resummation scheme, which are different from those shown in Fig.~\ref{Fig2}. The diagrams inside the pink box represents a sum of all ladder diagrams with the two-particle vertices.}
\label{Fig3}
\end{figure}

Before closing this section, we want to further comment on the diagram resummation scheme. It shall be noted that such a scheme is not unique. Beside the free choice on the local limit, it is another feature of our non-local expansion method that different diagram resummation schemes are allowed. 
For simplicity, in Eq.~(\ref{g-NL}) we only considered diagrams with one- and two-particle vertices. But there is no problem to include higher-order vertices in $\Lambda^{\omega}_{\alpha\beta}$. 
Other choices of the lower-order binding-block diagrams as well as different ways of constructing higher-order reducible diagrams are possible. The only constraint on the building-block diagrams is to require them to be 1PI, otherwise some diagrams will be duplicated after the resummation. 
As an example to illustrate such feasibility, we present another resummation scheme in Fig.~\ref{Fig3}.

Fig.~\ref{Fig3} contains only the one- and two-particle vertices, in this sense it is similar to Fig.~\ref{Fig2}. But it contains essentially different diagrams and represents a different resummation scheme. 
Here we sum the diagrams in Fig.~\ref{Fig3}(a) up to infinite order. One can find the corresponding expression for Fig.~\ref{Fig3}(a) in Eq~(\ref{G0}). 
Fig.~\ref{Fig3}(b) is same as Fig.~\ref{Fig2}(b), which is the lowest diagram concerning two-particle vertex. 
The diagram resummation in Fig.~\ref{Fig3}(c) is more subtle. Here we do not use the entire Fig.~\ref{Fig3}(c) as building-block diagram, but consider the sum of the local two-particle vertices in a ladder, which is graphically shown in the pink box in Fig.~\ref{Fig3}(c).
Mathematically, this sum is known as the Bethe-Salpeter equation.
\begin{eqnarray}\label{BSE}
&&\Gamma_{\alpha\beta}^{c/s}(12;34)=\gamma_{\alpha}^{c/s}(12;34)\delta_{\alpha\beta}  \nonumber \\
&&-T\sum_{\delta=1}^{N}\sum_{5,6}\gamma_{\alpha}^{c/s}(12;56)
\tilde{V}_{\delta\alpha}(\omega_{5})\tilde{V}_{\alpha\delta}(\omega_{6})
\Gamma_{\delta\beta}^{c/s}(65;34)\:.
\end{eqnarray}
From this equation, one can easily understand that, after summing all ladder diagrams, the resulting dressed two-particle vertex $\Gamma_{\alpha\beta}^{c/s}$ becomes non-local. 
As shown in Fig.~\ref{Fig3}(c), the dressed two-particle vertices from both the particle-hole horizontal and vertical channels are then used to construct the one-particle Green's function. 
A compact expression for Fig.~\ref{Fig3}(c) can be obtained by employing the crossing relation of $\Gamma_{\alpha\beta}^{c/s}$ in the horizontal and vertical channels, as well as the crossing symmetry of $\gamma_{\alpha}^{c/s}$. 
The expression for Fig.~\ref{Fig3} is in the end found to be:
\begin{equation}
\begin{aligned}
G_{\alpha\beta}^{\omega}&=\left[\frac{\mathbb{1}}{g^{0,-1}\mathbb{1}-V}\right]_{\alpha\beta}  \\
&-T\sum_{2}\gamma_{\alpha}^{\sigma_{1}\sigma_{1};\sigma_{2}\sigma_{2}}(11;22)\tilde{V}_{\alpha\alpha}(\omega_{2}) \\
&-\frac{T^{2}}{2}\sum_{\delta=1}^{N}\sum_{234}
\tilde{V}_{\beta\alpha}(\omega_{2})\tilde{V}_{\alpha\delta}(\omega_{3})\tilde{V}_{\delta\alpha}(\omega_{4}) \\
&\bigg\{ \gamma^{c}_{\alpha}(12;34)[\Gamma_{\delta\beta}^{c}(43;21) - \frac{1}{2}\gamma_{\beta}^{c}(43;21)] \\
&+3\gamma^{s}_{\alpha}(12;34)[\Gamma_{\delta\beta}^{s}(43;21)-\frac{1}{2}\gamma_{\beta}^{s}(43;21)]\bigg\}\:.
\end{aligned}
\end{equation}   

\section{Application: new insights into the dual-fermion
  approach}\label{dual-fermion} 
The non-local expansion scheme corresponds to the linked-cluster (or cumulant)
expansion approach~\cite{PhysRevB.43.8549, PhysRevB.67.075101, PhysRevE.89.063301, PhysRevB.67.075101, PhysRevLett.80.5389} if $V_{ij}$ is simply taken as $\mathbf{t}_{ij}$ in Eq.~(\ref{hamiltonian}) and all terms in Eq.~(\ref{G-expansion}) are considered (thus, no diagram resummation is required.).
The diagram-resummation algorithm in Eq.~(\ref{g-NL}) corresponds to the scheme discussed in Refs.~\onlinecite{Moskalenko-1, Moskalenko-2}. 
As discussed before, the choice of local system is quite flexible in the non-local expansion scheme.
In the following section, we will show that, by choosing a different local limit, the current scheme can be
connected to and further extend some well-known local many-body
approaches. 
As the first application, we will show that the DF approach exactly corresponds to one
special case of the non-local expansion scheme for a translationally
invariant system. 
From the current non-local expansion scheme, we want to explain the physics hidden 
behind the usage of the dual variables.

In Eq.~(\ref{G-expansion}), we have employed a direct expansion of
$V_{ij}$ and considered the expansion in orders.
Now, we want to perform the same expansion by introducing a dual
variable through the Hubbard-Stratonovich transformation. This is also
known as a strong-coupling expansion of the
Hubbard model~\cite{0022-3719-21-18-002, Pairault2000}.   
The main question we want to address in this section is that, by
taking a different form of $V_{ij}$, we can prove that the dual-fermion
approach~\cite{PhysRevB.77.033101} is equivalent to the current non-local
expansion scheme with the diagram resummation algorithm shown in
Eq.~(\ref{g-NL-k}).    

The DF approach is an elegant non-local extension of the
DMFT, which was proposed for systems with translational symmetry.
However, there is obvious no obstacle for it to apply to inhomogeneous
systems. One only needs to be careful with the breaking of
translational symmetry and formulate it in coordination, instead of
momentum space. 
Here, we would instead use Eq.~(\ref{g-NL-k}) and prove that the
current scheme can reproduce the DF approach for
a translationally invariant system, if the local action is taken as the
DMFT one.  

According to the DF approach, we add and subtract a local
dynamic function $\Delta_{i}(\omega)$ to Eq.~(\ref{full-action}),   
\begin{eqnarray}\label{dual-action}
  S&=&\sum_{i=1}^{N}{\cal S}_{i} + T\sum_{i,j}c_{i\omega}^{*}[{\cal
    G}^{-1}_{\sigma}(\omega)]_{ij}c_{j\omega} \nonumber\\
  &=&\sum_{i=1}^{N}{\cal S}_{i} - T\sum_{k,\omega}c_{k\omega}^{*}(
    \Delta_\omega-\epsilon_{k})c_{k\omega}\:,
\end{eqnarray} 
where ${\cal S}_{i}$ is the local action of the impurity at the $i$th
site, ${\cal S}_{i}=-(\omega+\mu) + \Delta_\omega + Un_{i\uparrow}n_{i\downarrow}$. 
$\Delta_{\omega}$, in principle, can be an arbitrary function. As
shown in the dual ferimon approach, diagrams for the DF
self-energy can be simplified if $\Delta_{\omega}$ is taken as the
hybridization function of the DMFT.
In Eq.~(\ref{dual-action}), $V_{k}^{\omega}$ is given as
$-(\Delta_\omega-\epsilon_{k})$, which contains
both local and non-local components.
One can still repeat the procedure shown in Sec.~\ref{formalism} to
calculate the non-local single-particle Green's function. 
Due to the inclusion of the local component in $V_{k}^\omega$, there
are additional diagrams which shall be considered.
Fortunately, such diagrams, after the renormalization of
$V_{k}^\omega$, have actually been included in Fig.~\ref{Fig2}.
Thus, the final expression of the single-particle Green's function
from the non-local expansion scheme Eq.~(\ref{g-NL-k}),
also applies to the current case. 

In the following, we want to show two important observations of the
DF approach. {\it First, we show that the physics origin of
the non-interacting DF propagator corresponds exactly to the
renormalized hybridization $\tilde{V}_{k}^{\omega}$.} To see this, we
need to slightly modify the construction of the DF approach. 
Different from the Hubbard-Stratonovich transformation generally
employed in the dual fermion approach~\cite{PhysRevB.77.033101}, we
set the coefficient of the mixing term between the original and the dual
variables to be one. 
The full action  in Eq.~(\ref{dual-action}) becomes
\begin{eqnarray}
  {\cal S}&=&{\cal S}[c^{*}, c;f^{*},f] =\sum_{i=1}^{N}{\cal S}_{i} -\nonumber\\
  && - T\sum_{k\omega}[f_{k\omega}^{*}c_{k\omega} + 
  c_{k\omega}^{*}f_{k\omega}
  +\frac{f_{k\omega}^{*}f_{k\omega}}{\Delta_{\omega}-\epsilon_{k}}]\:.  
\end{eqnarray}
Then we proceed in the same way as the normal DF approach~\cite{PhysRevB.77.033101} does to get the exact relation
between the lattice Green's function and the so-called DF
Green's function:
\begin{equation}\label{Gd}
G_{\alpha\beta}^{\omega} = (\Delta_{\omega}-\epsilon_{k})^{-1} + (\Delta_{\omega}-\epsilon_{k})^{-1}G_{k}^{d,\omega}(\Delta_{\omega}-\epsilon_{k})^{-1}\:,
\end{equation}
where $G_{k}^{d,\omega}$ is the single-particle propagator
defined for the dual variables, whose non-interacting part is given as:
\begin{equation}\label{Gd0}
G^{d,0,\omega}_{k} = -\left[ (\Delta_{\omega}-\epsilon_{k})^{-1}+g^{\omega}\right]^{-1}=\tilde{V}_{k}^{\omega}\:.
\end{equation}

From Eq.~(\ref{Gd0}), one immediately understands that the non-interacting dual
fermion propagator is not just a mathematical definition, it has clear
physical correspondence. It corresponds to the dressed hybridization
we obtained in Eq.~(\ref{dressed_hybridization}).
As shown in the DF approach~\cite{PhysRevB.77.033101}, the
coarse graining of $G^{d,0,\omega}_{k}$ corresponds to the DMFT
self-consistent equation.
\begin{equation}
-\sum_{k}\frac{1}{(\Delta_{\omega}-\epsilon_{k})^{-1}+g^{\omega}} = \sum_{k}\tilde{V}_{k}^{\omega}=0\:,
\end{equation}
which indicates that the average of the non-local
hybridization $\tilde{V}_{k}^{\omega}$ in momentum space is zero. 
This, from another perspective, reveals the local nature of the
DMFT, {\it i.e.} the coarse graining effect of the non-local fluctuations
vanishes in the single-particle level.

It also becomes transparent now that, in the non-interacting limit for the
dual variables, the DF approach can give rise to the
correct weak-coupling and the strong-coupling behaviors for a reason
exactly the same as we discussed in Eq.~(\ref{G0}). 
Furthermore, Eq.~(\ref{Gd0}) also implies that the self-energy in
the DF approach has the same set of Feynmann diagrams as the
Green's function in our non-local expansion scheme, {\it i.e.}
Fig.~\ref{Fig2}(b) and (c) are also the first two diagrams for the self-energy
in the DF approach.   
As the DF self-energy is known to be
analytic~\cite{PhysRevB.77.033101}, the lattice Green's function of the  
non-local expansion scheme is, therefore, also an analytic function.

What is more subtle about the DF approach, which is
the second important observation of this section, is that {\it it
  exactly the same as our non-local expansion scheme with the 
  diagram-resummation algorithm Eq.~(\ref{g-NL-k}).}
To prove this, we show the equivalence of Eq.~(\ref{Gd}) to 
Eq.~(\ref{g-NL-k}). Starting from Eq.~(\ref{Gd}), after substituting 
$G^{d,\omega}_{k}$ with $G^{d,0,\omega}_{k}$ and the DF
self-energy $\Sigma_{k}^{d,\omega}$, we have
\begin{eqnarray}\label{dual-NL}
G_{k}^{\omega} &=& (\Delta_{\omega}-\epsilon_{k})^{-1} +
\frac{(\tilde{V}_{k}^{\omega,
    -1}-\Sigma_{k}^{d,\omega})^{-1}}{(\Delta_{\omega}-\epsilon_{k})^{2}}\nonumber\\
&=&-\frac{1}{V_{k}^{\omega}} + \frac{(\tilde{V}_{k}^{\omega,
    -1}-\Sigma_{k}^{d,\omega})^{-1}}{(V_{k}^{\omega})^{2}}\nonumber\\
&=& -\frac{1}{V_{k}^{\omega}}
+\frac{1}{(V_{k}^{\omega})^{2}}\frac{V_{k}^{\omega}/(1-V_{k}^{\omega}g^{\omega})}{1-V_{k}^{\omega}\Sigma^{d,\omega}_{k}/(1-V_{k}^{\omega}g^{\omega})}\nonumber\\
&=&-\frac{1}{V_{k}^{\omega}}\left[1-\frac{1}{1-V_{k}^{\omega}(g^{\omega}+\Sigma^{d,\omega}_{k})}\right]\nonumber\\
&=&\frac{g^{\omega}+\Sigma^{d,\omega}_{k}}{1-V_{k}^{\omega}(g^{\omega}+\Sigma^{d,\omega}_{k})}\:.
\end{eqnarray} 
In the above derivation, Eq.~(\ref{dressed_hybridization}) and
Eq.~(\ref{Gd0}) have been employed.
As we discussed before, the first two DF self-energy function employs the
same Feynmann diagrams as shown in Fig.~\ref{Fig2}(b) and (c), {\it i.e.}
$g^{\omega}+\Sigma^{d,\omega}_{k}=\Lambda_{k}^{\omega}$. 
Thus, Eq.~(\ref{dual-NL}) is exactly the same as Eq.~(\ref{g-NL-k}). 
As no dual variable is involved in the construction of the non-local
expansion scheme, the dynamics and physical correspondence of each
term in Eq.~(\ref{g-NL-k}) is clear. The equivalence of the
DF and the non-local expansion scheme, thus, indicates that
the interacting nature of the lattice fermions in the DF
approach with the first two self-energy diagrams is given exactly by Fig.~\ref{Fig2}, in which three different
processes are taken as the most important contributions to the
dynamics at all length scales. 
The first one reproduces the non-interacting limit through the
non-local charge fluctuations; the non-local corrections to the
lattice self-energy are approximated by a single two-particle-scattering
mode in Fig.~\ref{Fig2}(b) and the coupling of two 
two-particle-scattering processes in Fig.~\ref{Fig2}(c).   
All three processes, then, are coupled through Eq.~(\ref{g-NL}) by the
non-local hybridization $V_{k}^{\omega}$. 
Such picture is less obvious in the DF approach.    

Based on the equivalence of  Eq.~(\ref{dual-NL}) and Eq.~(\ref{g-NL-k}), it is no surprise that the lattice self-energy of the non-local expansion scheme is same as that of the DF approach.
With Eq.~(\ref{g-NL-k}), the lattice self-energy is found to be:   
\begin{eqnarray}
\Sigma_{k}^{\omega} &=& \omega + \mu - \epsilon_{k} - 1/G_{k}^{\omega}
\nonumber\\
&=& \omega + \mu - \epsilon_{k} - /(g^{\omega} +
\Sigma_{k}^{d,\omega})+ V_{k}^{\omega}\nonumber\\
&=&\omega+\mu-\Delta_{\omega} - 1/(g^{\omega} + \Sigma_{k}^{d,\omega})\:.
\end{eqnarray}
Given the DMFT Dyson equation $g^{\omega, -1} = \omega+\mu -
\Delta_{\omega}-\Sigma_{loc}^{\omega}$, the above equation is then
simplified as 
\begin{eqnarray}\label{selfenergy}
\Sigma_{k}^{\omega} &=& \Sigma_{loc}^{\omega} + 1/g^{\omega} -
1/(g^{\omega} + \Sigma_{k}^{d,\omega}) \nonumber \\
 &=& \Sigma_{loc}^{\omega} +
\frac{\Sigma_{k}^{d,\omega}}{g^{\omega}(g^{\omega}+\Sigma_{k}^{d,\omega})}\nonumber\\
&=& \Sigma_{loc}^{\omega} +
\frac{\Sigma_{k}^{d,\omega}/(g^{\omega})^{2}}{1+g^{\omega}\Sigma_{k}^{d,\omega}/(g^{\omega})^{2}}\:.  
\end{eqnarray}
This is exactly same as Eq.~(18) of Ref.~\onlinecite{PhysRevB.79.045133}. 

\begin{figure}[htbp]
\centering
\includegraphics[width=\linewidth]{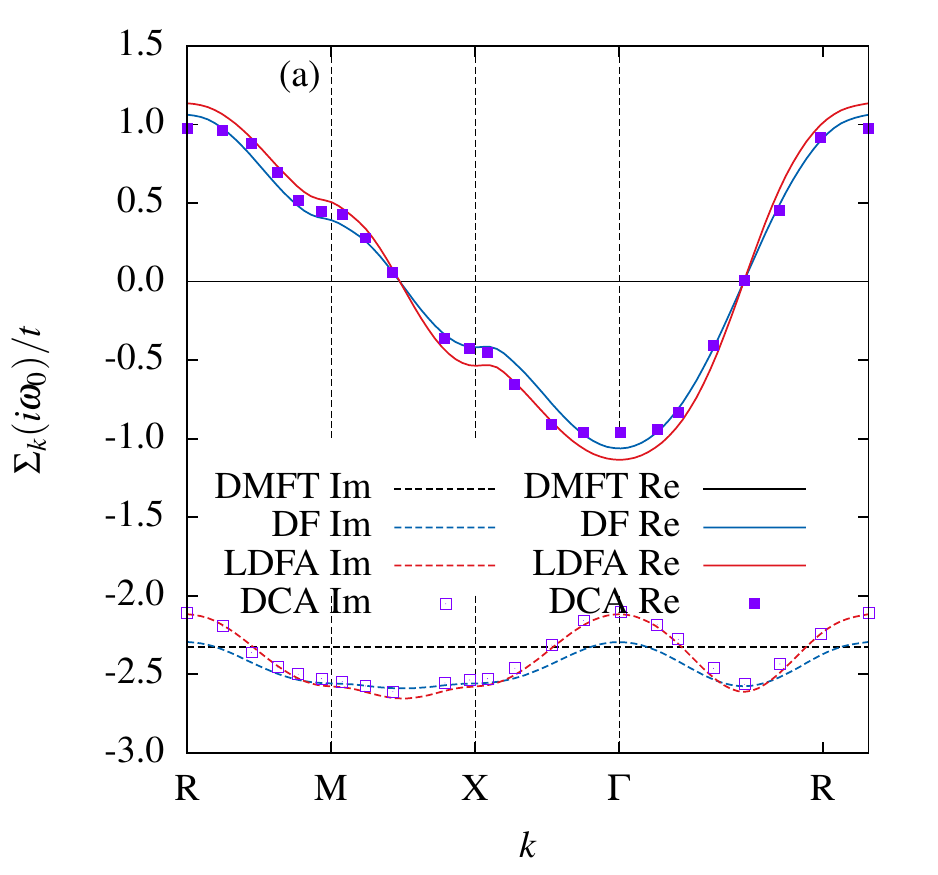}
\includegraphics[width=\linewidth]{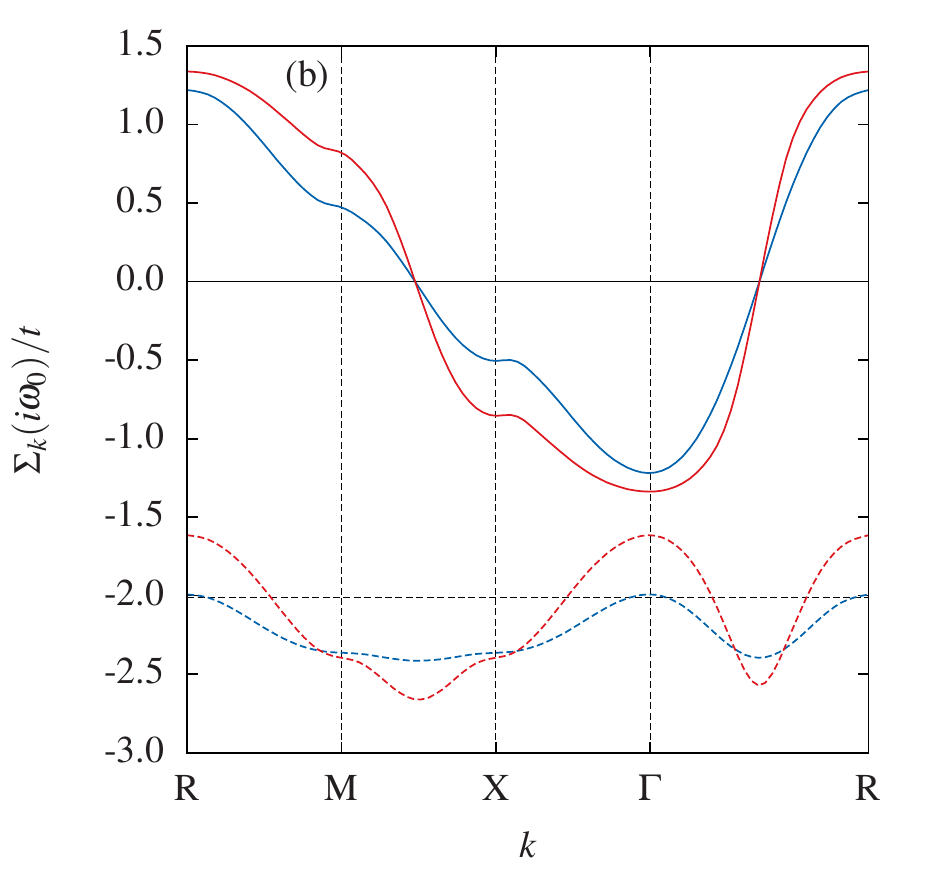}
\caption{The momentum dependence of the lattice self-energy in the
  non-local expansion scheme, as well as in the DF
  approach, for the single-band Hubbard model on a cubic lattice at
 (a) $\beta t=2$, (b) $\beta t=2.6$ and $U/t=8$. The solid and dashed lines correspond to
  the real and  imaginary parts of $\Sigma_{k}^{\omega}$ at the
  lowest Matsubara frequency. They agree well with the results from a
  calculation within DCA with a 100 site cluster~\cite{PhysRevB.88.115101} which are replotted as squares here. } 
\label{Fig4}
\end{figure}
In the last part of this section, we present a calculation for a
correlated system with translational symmetry by using the non-local
expansion scheme Eq.~(\ref{g-NL-k}), which also represents a solution of the DF
approach.
In order to fully demonstrate the non-locality generated by these two
approaches, we show, in Fig.~\ref{Fig4}, the lattice self-energy for a
three dimensional cubic lattice with $U/t=8$ along the high symmetry line R-M-X-$\Gamma$-R.
Results from large cluster DCA calculations~\cite{PhysRevB.83.075122} with the same parameters are available for comparison. 
The DMFT solutions have no momentum dependence, thus, they are straight lines in Fig.~\ref{Fig4}. The inclusion of non-locality, as shown in Eq.~(\ref{g-NL-k}), generates dispersion in the real- and imaginary-part of the lattice self-energy around the DMFT solutions. 
Here, results are shown for two different temperatures. The lower temperature ($\beta t=2.6$, see Fig.~{Fig4}(b)) results display more pronounced momentum dependence of $\Sigma_{k}(i\omega_{0})$ than the higher temperature ($\beta t=2.0$, see Fig.~\ref{Fig4}(b)) ones. With the decrease of temperature, the spatial fluctuations are enhanced. 
In addition to the results from the DF approach and the non-local expansion scheme, Fig.~\ref{Fig4} also shows the corresponding solution from Eq.~(\ref{g-NL-k}) with a modified $\Lambda_{k}^{\omega}$. In addition to the diagrams shown in Fig.~\ref{Fig2}, all ladder-type diagrams in the both horizontal and vertical channels are included in this calculation. The additional scattering processes included in the ladder diagrams yields a better inclusion of non-locality, thus, a more pronounced momentum dependence of $\Sigma_{k}(i\omega)$ can be observed. Without overhead thinking, one can immediately understand that the adoption of the ladder diagrams in $\Lambda_{k}^{\omega}$ (, which can be generated by replacing $\gamma_{\beta}$ in Fig.~\ref{Fig2} by the corresponding ladders $\Gamma_{\delta\beta}$) in the non-local expansion scheme corresponds exactly to the ladder dual fermion approach (LDFA)~\cite{PhysRevLett.102.206401}. We want to note that the LDFA results show surprisingly good agreement (especially the imaginary part) with the one from a DCA calculations with a cluster of 100 sites (see Fig. 5 in Ref.~\onlinecite{PhysRevB.83.075122}).
Compared to the time-consuming large cluster DCA calculations, the present non-local expansion scheme, as well as the DF approach, is numerically much more economical.  
Note, in the DCA$^{+}$ scheme~\cite{PhysRevB.88.115101} proposed recently, a continuous lattice self-energy can also be achieved with much less numerical effort, which represents another promising route to understand the non-locality of a quantum many-body system. 

\section{Outlook}
Besides the intuitive understanding of the DF approach, in this section, we want to briefly outlook other possible applications of the non-local expansion scheme. More detailed discussions and the corresponding results will be presented elsewhere. 

{\it First, we want to show that the non-local expansion scheme can be used as a quick cluster solver for the Cellular-DMFT. } In the Cellular-DMFT, the translational symmetry is naturally broken due to the different intra- and inter-cluster hoping amplitudes.  Though in some special cases (for example in a 2x2 square cluster), the cluster momenta are still a good quantum number, we usually have to work with the situation that translational invariance is lost. Thus, Eq.~(\ref{g-NL}) and (\ref{g-NL-L1}) shall be used. In the Cellular-DMFT, the Weiss field contains the inter- and intra-cluster components. For a calculation with the non-local expansion scheme, in the first step we only need the diagonal component of the hybridization function $\Delta_{i,i}^{\omega}$, which we adopt to construct the local Weiss field ${\cal G}_{i}^{\omega}=-1/(\omega+\mu-\Delta_{i,i}^{\omega})$, 
which is different from the diagonal component of the Weiss field from the cellular-DMFT ${\cal G}_{i,i}^{\omega}=-[(\omega+\mu)\mathbb{1}-\Delta]^{-1}_{i,i}$. 
Thus, the separation of the local and non-local degrees of freedom looks like:
\begin{equation}
{\cal S} = \sum_{i=1}^{N}{\cal S}_{i} - T\sum_{i\ne j}\sum_{\omega}c_{i\omega\sigma}^{*}\Delta_{i,j}^{\omega}c_{j\omega\sigma}\:.
\end{equation}
With the new ${\cal G}_{i}^{\omega}$, an interacting impurity problem defined by ${\cal S}_{i}$ is then solved to get the local self-energy $\Sigma_{i}^{\omega}$ and the local single-particle propagator $g_{i}^{\omega}$, as well as the connected four-point correlation function $\gamma_{i}^{s/c}$.
In the second step, the non-local expansion scheme Eq.~(\ref{g-NL}) will be issued to calculate both the local and non-local components of the single-particle propagator $g_{ij}^{\omega}$ and the self-energy function $\Sigma_{i,j}^{\omega}$. The new Weiss field then can be calculated from $\Sigma_{i,j}^{\omega}$, which closes the Cellular-DMFT self-consistency loop. 
As one can immediately understand, the application of the non-local expansion scheme to the Cellular-DMFT equations can reduce the computational effort of solving a cluster problem to that of solving an impurity problem. Thus, a larger cluster calculation can be expected for the Cellular-DMFT with the non-local expansion scheme as cluster solver.

{\it Second, the non-local expansion scheme can extend the R-DMFT to incorporate non-locality in the self-energy function.}
R-DMFT is widely used in the study of disordered correlated systems~\cite{PhysRevLett.78.3943, PhysRevB.77.054202, PhysRevLett.91.066603, 0034-4885-68-10-R02, PhysRevLett.94.187203}. 
The disorders we considered here can be the different site energies $\epsilon_{i}$ and electronic correlations $U_{i}$, which are local in space.  Thus, a DMFT-fashion calculation can be formulated.  
The R-DMFT iteration cycle begins with the calculation of the lattice Green's function, 
\begin{equation}\label{R-DMFT}
G^{\omega}_{i,j} = \left[\frac{1}{(\omega+\mu)\mathbb{1} - \hat{t} - \hat{\epsilon} - \Sigma^{\omega}}\right]_{i,j}\:,
\end{equation}
in which $\hat{t}$ and $\hat{\epsilon}$ are of matrix forms. Matrix $\Sigma^{\omega}$, in R-DMFT, is simplified to have only diagonal elements $\Sigma^{\omega}_{i}$.  
For each site $i$, one defines a Weiss field ${\cal G}^{\omega}_{i}=[G_{i,i}^{\omega, -1} + \Sigma_{i}^{\omega}]^{-1}=(\omega+\mu)-\epsilon_{i}-\Delta_{i}^{\omega}$, from which the local Green's function $g_{i}^{\omega}$ and a new local self-energy $\Sigma_{i}^{\omega}$ can be calculated in the conventional DMFT.  
Normally, the R-DMFT closes the iteration cycle by substituting the new $\Sigma_{i}^{\omega}$
to Eq.~(\ref{R-DMFT}). 
Now, with the non-local expansion scheme, a non-local self-energy can be obtained for the R-DMFT. 
Like in the DF approach, we separate the local and non-local degrees of freedom by adding and subtracting a local dynamic function $\Delta_{i}^{\omega}$ to the denominator of Eq.~(\ref{R-DMFT}). The non-local hybridization used for expansion in the non-local expansion scheme can be easily identified as $-(\Delta_{i}^{\omega}\mathbb{1}-\hat{t})$. With Eq.~(\ref{g-NL}), a self-energy matrix with off-diagonal elements can be obtained, which generates the non-trivial spatial fluctuations to the R-DMFT. Here, it has to be noted that at each site $i$, $g_{i}^{\omega}$ and $\gamma_{i}^{c/s}$ take different values due to the site-dependent disorders. Thus, the single-particle charge fluctuations and the two-particle scattering modes differ at different sites. For this reason, the non-local expansion scheme essentially extends the non-local correlations of the R-DMFT from the one-particle to the two-particle level. 

\section{Conclusions}\label{conclusion}
In this paper, we proposed a non-local expansion scheme to study correlated many-fermion systems. By separating the local and non-local degrees of freedom and assuming that the local system can be solved exactly (which, at least, can be achieved numerically), we can treat the non-local terms as a small perturbation to the local degrees of freedom. A non-local expansion around the solution of the local system can generate, order by order, the non-local corrections to the local solutions. This scheme can be widely applied to correlated systems with/without translational symmetry. Numerically, it is as economical as the DMFT, thus, a continuous momentum dependence in the self-energy function can be achieved easily in this scheme. 
As the first application, we have proven that the DF approach can be beautifully explained as one special case of the non-local expansion scheme presented in this work. When the local system is taken as the DMFT impurity and $\Lambda_{k}^{\omega}$ is approximated as the diagrams shown in Fig.~\ref{Fig2} or ladder-type diagrams, the DF or the LDFA can be reproduced. 

\acknowledgments
We thank A.N. Rubtsov, H. Lee, H. Monien, H. Hafermann, A.I. Lichtenstein and W. Hanke for the fruitful collaborations on the dual-fermion approach. We thank A. Fleszar, K.S. Chen and Tin Ribic for reading the manuscript.  
We acknowledge the support from the DFG Grants No. Ha 1537/23-1 within
the Forschergruppe FOR~1162 and SPP Ha 1537/24-2.

\bibliography{ref}
\end{document}